%% file: main.tex
\documentclass[a4paper]{scrarticle}
\pdfoutput=1

\usepackage[utf8]{inputenc} % allow utf-8 input
\usepackage[T1]{fontenc}    % use 8-bit T1 fonts
\usepackage{float}
\usepackage{hyperref}       % hyperlinks
\usepackage{booktabs}       % professional-quality tables
\usepackage{amsmath,amsfonts,amssymb}       % blackboard math symbols
\usepackage{nicefrac}       % compact symbols for 1/2, etc.
\usepackage{microtype}      % microtypography
\usepackage{listings,xcolor}
\input{lstinit}

\floatstyle{plaintop}
\newfloat{lstfloat}{tbp}{lop}
\floatname{lstfloat}{Listing}

\usepackage{graphicx}
\usepackage{tikz,xcolor}
\usetikzlibrary{matrix,calc,fadings}

\usepackage{xurl}
\urldef\urldefCPythonGitHub\url{https://github.com/python/cpython}
\urldef\urldefCPythonHomepage\url{https://www.python.org/}
\urldef\urldefNumPyGitHub\url{https://github.com/numpy/numpy}
\urldef\urldefNumPyHomepage\url{https://numpy.org/}
\usepackage{siunitx}
\DeclareSIUnit\byte{B}
\usepackage{pgfplots}
\pgfplotsset{compat=1.11}
\newenvironment{customlegend}[1][]{%
    \begingroup
    \csname pgfplots@init@cleared@structures\endcsname
    \pgfplotsset{#1}%
}{%
    \csname pgfplots@createlegend\endcsname
    \endgroup
}%
\def\addlegendimage{\csname pgfplots@addlegendimage\endcsname}

\newcommand{\addlegendimageintext}[1]{%
    \tikz {
        \begin{customlegend}[
            legend entries={\empty},
            legend style={
                draw=none,
                inner sep=0pt,
                column sep=0pt,
                nodes={inner sep=0pt}}]
        \addlegendimage{#1}
        \end{customlegend}
    }%
}

\title{Reverse-Mode Automatic Differentiation \\ of Compiled Programs}

\author{Max Aehle, Johannes Blühdorn, Max Sagebaum, \\ and Nicolas R.\ Gauger}
\date{}

\begin{document}
\newcommand{\RR}{\mathbb R}
\newcommand{\Cpp}{C{\ttfamily++}}

\maketitle

\begin{abstract}
Tools for algorithmic differentiation (AD) provide accurate derivatives of computer-implemented functions for use in, e.\,g., optimization and machine learning (ML). However, they often require the source code of the function to be available in a restricted set of programming languages. As a step towards making AD accessible for code bases with cross-language or closed-source components, we recently presented the forward-mode AD tool Derivgrind. It inserts forward-mode AD logic into the machine code of a compiled program using the Valgrind dynamic binary instrumentation framework.

This work extends Derivgrind, adding the capability to record the real-arithmetic evaluation tree, and thus enabling operator overloading style reverse-mode AD for compiled programs.  
We maintain the high level of correctness reported for Derivgrind's forward mode, failing the same few testcases in an extensive test suite for the same well-understood reasons. 
Runtime-wise, the recording slows down the execution of a compiled 64-bit benchmark program by a factor of about 180.
\end{abstract}

\section{Introduction}\label{sec:intro}
Gradient-based numerical optimization algorithms have become an indispensable tool to improve engineering designs or train machine learning (ML) models, besides many other applications. Such algorithms rely on the gradient vector $\nabla f(x^{(t)}) \in \RR^n$ of a computer-implemented objective function $f: \RR^n \to \RR$ to iteratively step from one set of, e.\,g., design or ML model parameters $x^{(t)} \in \RR^n$ to the next iterate $x^{(t+1)}\in\RR^n$, such that the engineering performance is maximized or the training error is minimized in the course of many iterations. While $\nabla f(x^{(t)})$ could be found numerically using difference quotients, this would require ${\mathcal O}(n)$ evaluations of $f$ to approximate all $n$ partial derivatives, and incur numerical errors. In contrast, the \emph{reverse mode} of \emph{algorithmic differentiation} (AD), or \emph{backpropagation} in ML, requires a run-time proportional to the evaluation of $f$ to compute all components of $\nabla f(x^{(t)})$ at once, up to floating-point precision. 

The key idea of reverse-mode AD is to represent the computer-implemented function as a sequence of simple real-arithmetic calculations, with well-known differentiation rules, acting on a set of variables. Every variable $a$ appearing in the implementation of $f$ is matched by a \emph{adjoint} or \emph{bar value} $\bar a$, which tracks the derivative of the output variable $f(x)$ with respect to (w.\,r.\,t.)\ the value of $a$ at a certain ``point of time'' during the evaluation of $f$.  If this ``point of time'' is \emph{after the last real-arithmetic calculation}, the bar value of the output variable itself is~$1$ and all other bar values are~$0$, and this is how the bar values are initialized. On the other hand, the derivatives w.\,r.\,t.\ the values of input variables \emph{before the first real-arithmetic calculation} form the sought gradient $\nabla f(x)$. In order to propagate the bar values backwards, reverse-mode AD iterates, \emph{in reverse order}, throught the real-arithmetic calculations performed by $f$, matching every step $a = \phi(a_1, \dots, a_k)$ by updates
\begin{align}
\bar a_i\,+\!\!&= \frac{\partial \phi}{\partial a_i}(a_1,\dots,a_k) \cdot \bar a \qquad \text{for $i=1,\dots,k$, and} \label{eq:adjoint-update} \\
\bar a &= 0. \label{eq:adjoint-zero}
\end{align}
The increment in \eqref{eq:adjoint-update} reflects that the derivative of the output variable w.\,r.\,t.\ the value of $a_i$ \emph{after} the calculation differs from the derivative w.\,r.\,t.\ the value of $a_i$ \emph{before} the calculation by the implicit dependency of the output variable on the variable $a$ that itself depends on $a_i$. Equation~\eqref{eq:adjoint-zero} refers to the fact that the overwritten value of $a$ \emph{before} the assignment has no influence on the output variable.

In advanced use cases for functions $f: \RR^n \to \RR^m$ with multiple output variables $f_1(x)$, \dots, $f_m(x)$ whose bar values are initialized with $\bar f_1, \dots, \bar f_m$, $\bar a$ represents a linear combination $\tfrac{\partial f_1}{\partial a}\cdot \bar f_1 + \dots + \tfrac{\partial f_m}{\partial a} \cdot \bar f_m$ of the partial derivatives of the output variables w.\,r.\,t.\ the values of $a$.

Two main strategies for implementing \eqref{eq:adjoint-update} and \eqref{eq:adjoint-zero} in a reverse iteration through the real-arithmetic calculations performed by $f$ have been reported in the literature. \emph{Source transformation} tools, in a general sense, produce new code containing \eqref{eq:adjoint-update} and \eqref{eq:adjoint-zero} in a reversed control flow. 
Besides the memory needed to store the bar values, they allocate a stack-like data structure called the \emph{tape}. It stores the values of variables before they are reassigned, if needed to compute the partial derivatives in \eqref{eq:adjoint-update}, and additional information to reconstruct the control flow. 
As an example for a source transformation tool, TAPENADE \cite{hascoet_tapenade_2013} differentiates source code written in Fortran and C.
Enzyme \cite{NEURIPS2020_9332c513} operates on the intermediate representation (IR) of the LLVM compiler infrastructure instead, and is thus applicable to software projects written in any mix of languages for which LLVM front-ends exist. These projects may even involve static libraries if their LLVM IR bitcode is available. 

\emph{Operator-overloading} AD tools like ADOL-C \cite{Walther2012Gsw}, CoDiPack \cite{SaAlGauTOMS2019}, the autograd tool \cite{maclaurin2015autograd} used by PyTorch \cite{NEURIPS2019_9015}, and the internal AD tool of TensorFlow \cite{tensorflow2015-whitepaper}, use the tape to store information about the entire real-arithmetic evaluation tree instead of producing code. They can thus avoid parsing and transforming control structures in the implementation of $f$, and only need to receive a stream of all the real-arithmetic operations performed while evaluating $f$. The most convenient way of producing this stream is to provide a special floating-point type replacing, e.\,g., the C \lstinline[language={}]|double| or Python \lstinline[language=python]|np.array| types. Apart from the floating-point value, this new type contains an \emph{index} to identify the variable, and overloads the arithmetic operators and math functions with index handling and tape recording logic. Typically, such tools are designed for one particular programming language and the new type must be used in large parts of the source code.

All of these approaches restrict the set of supported programming languages. For example, we are not aware of any AD tool that could easily differentiate a Python function involving Python C extension modules, so currently a lot of scientific software written in \Cpp{} needs rewriting in order to be included in a ML framework. Due to this language barrier, many potentially interesting  studies about applications of optimization or ML across science and engineering are practically infeasible. If parts of the real arithmetic are only available in compiled form, these approaches become outright inapplicable. Furthermore, the necessary efforts to integrate the AD tool with the source code of $f$ might render exploratorive studies uneconomical.

Analogous considerations regarding \emph{foward-mode AD} have led us to the development of the machine code based forward-mode AD tool Derivgrind \cite{aehle_forward-mode_2022}. Forward-mode AD is a related technique, which keeps track of the \emph{dot value} $\dot a$ for each variable $a$, defined as its derivative with respect to a single input variable, matching every calculation $a = \phi(a_1, \dots, a_k)$ by an update
\begin{equation}\label{eq:dot-update}
\dot a = \frac{\partial \phi}{\partial a_1}(a_1,\dots,a_k) \cdot \dot a_1 + \dots + \frac{\partial \phi}{\partial a_k}(a_1,\dots,a_k) \cdot \dot a_k.
\end{equation}
Derivgrind, in the version presented in \cite{aehle_forward-mode_2022}, inserts the forward-mode AD logic \eqref{eq:dot-update} into compiled programs running on Linux systems on the 32-bit x86 and the 64-bit amd64 architectures, using the dynamic binary instrumentation framework Valgrind \cite{valgrind-paper,valgrind-doc}. In this work, we extend Derivgrind by the feature to record the real-arithmetic evaluation tree of a compiled program. In combination with a simple tape evaluation program that reads the tape backwards and performs \eqref{eq:adjoint-update} and \eqref{eq:adjoint-zero}, we thus unlock operator overloading style reverse-mode AD for software projects involving cross-language or undisclosed sources, requiring minimal integration efforts.  %As the procedures to identify the performed real arithmetic are precisely the same for both modes of AD, we focus on the differences between the existing forward-mode AD (\emph{Dotgrind}) and the new tape recording pass (\emph{Bargrind}) logic and usage. 

Specifically, this work makes the following contributions.
In Section~\ref{sec:instrumentation-code}, we describe the index handling and tape recording logic that our extended version of Derivgrind inserts into the compiled program. While this process only requires access to the machine code, Derivgrind provides two interfaces to specify input and output variables by means of source code, as described in Section~\ref{sec:io-varname}. In Section~\ref{sec:io-funarg}, we present two \emph{external function} interfaces that allow ML researchers to conveniently use derivatives of compiled functions in dynamic libraries from a ML framework. Section~\ref{sec:evaluation} contains a verification study involving a broad collection of regression tests, and a performance study based on a finite-difference solver for the Burgers' partial differential equation (PDE). We close with a summary and outlook in Section~\ref{sec:conclusion}.

\section{Machine Code Instrumentation to Record the Tape}\label{sec:instrumentation-code}

\subsection{Valgrind and VEX IR} 
Derivgrind has been implemented within the Valgrind framework for building dynamic binary instrumentation tools \cite{valgrind-paper,valgrind-doc}. The Valgrind core loads the machine code of the so-called \emph{client program} $f$, including dynamically linked and loaded libraries, and presents portions of it to Derivgrind using the mostly architecture-independent \emph{VEX} intermediate representation. 

VEX code is a list of \emph{statements}, some of which move data between memory, registers of a synthetic CPU, and additional temporaries. There is a statement type for jumps, and special \emph{dirty call} statements allow to execute arbitrary C functions. Many arguments to statements, like memory addresses or the data to be written, are specified by VEX \emph{expressions} that acquire their value when the statement is run. Expressions can, e.\,g., load data from temporaries, registers and memory; evaluate integer arithmetic, floating-point arithmetic, and logical operations and comparisons; evaluate if-then-else constructs; call C functions without side effects; or simply represent a constant value. 

Derivgrind \emph{instruments} the presented portions of VEX code, inserting the index-handling and recording logic described in this section. Afterwards, the Valgrind core runs the instrumented code on the synthetic CPU.

\subsection{Index Handling}
\paragraph{Choice of Index Management} We identify every potential floating-point variable in the program by an 8-byte unsigned integer \emph{index}. \emph{Passive} variables, which do not depend on any input variable, share the default index~0. \emph{Active} variables, either declared inputs or the result of a real arithmetic operation with at least one active operand, are consecutively assigned the indices 1, 2, 3, \dots. We allow copies of the same active variable to share its index, which makes the AD instrumentation of data moves very easy. This approach is known as \emph{linear index management} with \emph{assign optimization} \cite{sagebaum2021index}. We cannot reuse indices after their last copy has disappeared, as we have no robust and efficient means to recognize this condition. 

\paragraph{Shadow Storage for Indices} Data in VEX can be stored in temporaries, registers of the CPU, and memory. As in the forward mode, we create additional ``shadow'' storage locations for any of these to store data for the AD instrumentation.
VEX identifies temporaries and registers by non-negative integers, and we acquire their respective shadows by adding multiples of suitable offsets $m_\text{tmp}$ and $m_\text{gs}$, respectively. 
Our shadow memory tool, implemented according to \cite{shadowmem-nethercote}, acts like an associative array mapping memory addresses to AD data. It enables Derivgrind to store shadow data in addition to, and without interfering with, the original data at the same address. 

In the forward mode, we used a single layer of shadow storage to store dot values on top of floating-point data in the same binary floating-point format. In the reverse mode, the shadow memory layout is more complicated because we need to store 8-byte indices on top of 4-byte floating-point values. Constrained by the fact that Valgrind provides only twice as many shadow registers as there are original registers, we decided to use two layers of shadow memory in the following way. No matter whether a floating-point value at some storage location is encoded in the 4- or 8-byte IEEE~754 formats \lstinline[language=ieee754]+binary32+ or \lstinline[language=ieee754]+binary64+ \cite{ieee754}, or in the 10-byte x87 double-extended precision format, we use the first four bytes of the ``lower'' and ``higher'' layer to store the less- and more-significant four-byte halves of the eight-byte index, respectively. The remaining 0, 4 or 6 bytes in either shadow layer are unused.

\paragraph{Indices of AD Inputs and Outputs}
In Section~\ref{sec:identifying-inputs-outputs}, we describe \emph{client requests}, which are a way for the client program to specify  memory locations as AD input and output variables. They trigger the following actions of Derivgrind:
\begin{itemize}
\item In order to declare a variable $x$ as an AD input, Derivgrind acquires a new index by recording the equivalent of a statement $x=0$, writes the upper and lower halves of it into the respective shadow memory locations for $x$, and appends the index to an \emph{input index file}.
\item In order to declare a variable $y$ as an AD output, Derivgrind records the equivalent of a statement $y_\text{copy} = 1.0\cdot y$ with an auxiliary variable $y_\text{copy}$, and appends the new index assigned to $y_\text{copy}$ to an \emph{output index file}. Acquiring an extra index in this way prevents bugs in the case that several copies of one variable, which share the same index, are individually declared as an AD outputs.
\end{itemize}

\paragraph{Instrumentation of Data Move Statements and Expressions} 
For any source and target locations of a copy operation, the respective two layers of shadow storage contain the entire AD information on the stored data, in a universal and translation-invariant format. Thus, the inserted AD logic simply moves the content of both shadow layers in exactly the same way as the original data. This type-agnostic rule is analogous to the forward mode, and it comes with the same exception of atomic compare-and-swap (CAS) statements. CAS moves depend on whether a shared state has been modified, and their instrumentation must extend this check to the shadow of the shared state, as discussed in mory detail in \cite{aehle_forward-mode_2022}.

\subsection{Tape Recording}
\paragraph{Tape Layout}
Derivgrind's tape is organized as a sequence of 32-byte blocks and there is one block for each assigned index. 
If the index $i$ was assigned to the result $a$  of a binary real arithmetic operation $a = \phi(a_1,a_2)$, the $i$-th block stores the 8-byte indices of $a_1$ and $a_2$, as well as the partial derivatives $\tfrac{\partial \phi}{\partial a_1}(a_1,a_2)$, $\tfrac{\partial \phi}{\partial a_2}(a_1, a_2)$ as 8-byte \lstinline[language=ieee754]|binary64|s.
Storing the partial derivatives is known as \emph{Jacobian taping} \cite{SaAlGauTOMS2019}.
For the default index $i=0$ and any index assigned to a declared input variable, we mandate that the two indices and partial derivatives in the corresponding block are \lstinline+0+ and \lstinline+0.0+, respectively. 
Figure~\ref{fig:vex-multiplicationsimple} displays the state of the tape and the index files after two variables $x_1=3$, $x_2=-4$ were declared as AD inputs, then multiplied, and the result was declared as an AD output.

\newcommand{\tapeblock}[4]{ $\begin{smallmatrix} \text{#1} & \text{#2} \\ \text{#3} & \text{#4} \end{smallmatrix} $}

\begin{figure}
\centering
\begin{tikzpicture}

\matrix (tape) at (0,-7.9) [matrix of nodes,nodes={rectangle,draw=black,font=\ttfamily\color{green!40!black},minimum width=6cm,minimum height=0.7cm,anchor=center},column sep=-\pgflinewidth,minimum height=0.4cm] { \tapeblock{00000000\,00000000~~}{~~00000000\,00000000}{$0.0$}{$0.0$} \\ \tapeblock{00000000\,00000000~~}{~~00000000\,00000000}{$0.0$}{$0.0$} \\ \tapeblock{00000000\,00000000~~}{~~00000000\,00000000}{$0.0$}{$0.0$} \\ \tapeblock{00000000\,00000001~~}{~~00000000\,00000002}{$-4.0$}{$3.0$} \\ \tapeblock{00000000\,00000003~~}{~~00000000\,00000000}{$1.0$}{$0.0$} \\ };
\draw ($(tape-1-1.west)+(-2.8,0)$) node [anchor=west,font=\ttfamily\scriptsize,align=left] {00000000\,00000000};
\draw ($(tape-2-1.west)+(-2.8,0)$) node [anchor=west,font=\ttfamily\scriptsize,align=left] {00000000\,00000001};
\draw ($(tape-3-1.west)+(-2.8,0)$) node [anchor=west,font=\ttfamily\scriptsize,align=left] {00000000\,00000002};
\draw ($(tape-4-1.west)+(-2.8,0)$) node [anchor=west,font=\ttfamily\scriptsize,align=left] {00000000\,00000003};
\draw ($(tape-5-1.west)+(-2.8,0)$) node [anchor=west,font=\ttfamily\scriptsize,align=left] {00000000\,00000004};
\draw (tape-1-1.north west) node [anchor=south west] (recordedtapelabel) {\itshape \!\!Recorded tape:};

\draw[very thick,<-] ($(tape-5-1.south west)+(-3,0)$) -- ($(tape-1-1.north west)+(-3,0)$) node[midway,above,sloped,font=\scriptsize] {index of block on tape};

\draw[<-,shorten <=0.1cm,shorten >=0.1cm] (tape-1-1.east) -- ($(tape-1-1.east)+(1,0)$) node[right,font=\scriptsize,align=left] {dummy block for \\ passive index \lstinline+0+};
\draw[<-,shorten <=0.1cm,shorten >=0.1cm] (tape-2-1.east) -- ($0.5*(tape-2-1.east) + 0.5*(tape-3-1.east) + (1,0)$) node[right,font=\scriptsize,align=left] (inputlabel) {declaration of \\ AD inputs};
\draw[<-,shorten <=0.1cm,shorten >=0.1cm] (tape-3-1.east) -- (inputlabel.west);
\draw[<-,shorten <=0.1cm,shorten >=0.1cm] (tape-4-1.east) -- ($(tape-4-1.east)+(1,0)$) node[right,font=\scriptsize,align=left] {differentiation rule \\ for multiplication};
\draw[<-,shorten <=0.1cm,shorten >=0.1cm] (tape-5-1.east) -- ($(tape-5-1.east)+(1,0)  + 0.5*(tape-5-1.east)-0.5*(tape-4-1.east)$) node[right,font=\scriptsize,align=left] {declaration of \\ AD output};

\draw ($(tape-5-1.south west)+(0,-1)$) node [anchor=north west,font=\ttfamily\scriptsize\color{green!40!black},align=left,rectangle,minimum width=1.1cm,draw=black] (inputfile) {00000000\,00000001 \\ 00000000\,00000002 \\ ~};
\draw (inputfile.west) node[anchor=east] (inputfilelabel) {\itshape \!\!Input indices:~~};
\draw (inputfile.east) node[anchor=west] (outputfilelabel) {\quad \itshape Output indices:~~};
\draw (outputfilelabel.east) node [anchor=west,font=\ttfamily\scriptsize\color{green!40!black},align=left,rectangle,minimum width=1.1cm,draw=black] (outputfile) {00000000\,00000004\\ ~ \\ ~ };
\end{tikzpicture}
\caption{Tape and index files after declaring two variables $x_1=3$, $x_2=-4$ as AD inputs, multiplying them, and declaring the result $y=x_1 \cdot x_2$ as an AD output. Each of the five tape blocks stores two indices (as 8-byte unsigned integers) and two partial derivatives (as 8-byte \lstinline|binary64|s, denoted here as decimal floating-point numbers).}
\label{fig:vex-multiplicationsimple}
\end{figure}
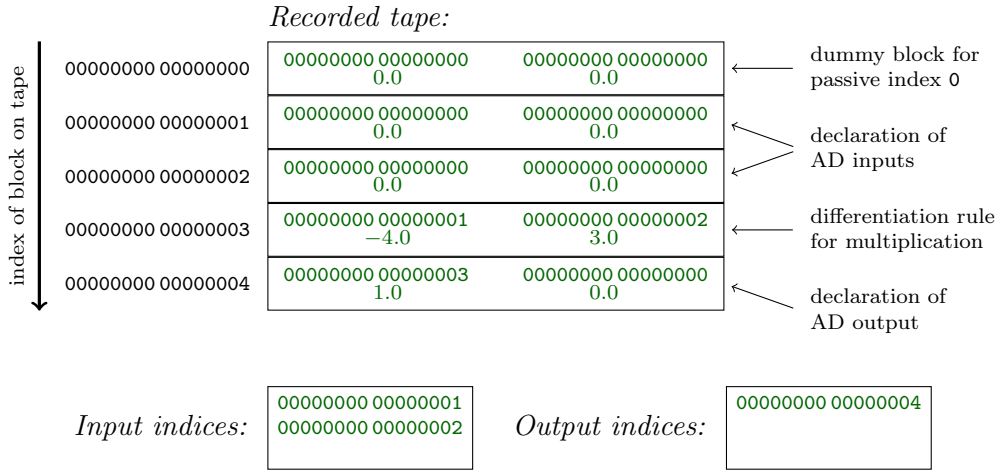

\paragraph{Instrumentation of Floating-Point Operations, Special Case} For a scalar floating-point VEX operation with two operands, the recording mode of Derivgrind complies with the previous paragraph by inserting the following instrumentation. First, the index of each differentiable operand is assembled from the two 4-byte halves, and the respective partial derivatives are computed according to the suitable differentiation rule. Then, a dirty call statement invokes C code that appends the new block to the tape and returns the number of blocks on the tape. This number is a new index representing the result of the VEX operation, so finally, VEX expressions for its upper and lower halves are generated. In the special case that both operands have index \lstinline+0+, both are passive and therefore the result is passive, thus no block is written and the index \lstinline+0+ is used for the result of the operation.

\paragraph{Instrumentation of Floating-Point Operation, General Case}
The recording-mode instrumentation of VEX operations with a single floating-point operand follows the same procedure, using the index \lstinline+0+ and the partial derivative \lstinline+0.0+ for the second operand. Operations $a = \phi(a_1, a_2, a_3)$ with three floating-point operands, like a fused multiply-add operation, are represented by two consecutive blocks. The first block stores the indices of $a_1$ and $a_2$ along with the partial derivatives $\tfrac{\partial \phi}{\partial a_1}$ and $\tfrac{\partial \phi}{\partial a_2}$. The second block stores the indices of this intermediate block and $a_3$ along with the partial derivatives \lstinline+1.0+ and $\tfrac{\partial \phi}{\partial a_3}$, respectively. 

Single-instruction-multiple-data (SIMD) instructions are decomposed into their componentwise, scalar operations. 
Figure~\ref{fig:vex-multiplication-instrumented} demonstrates the actions of Derivgrind's recording-mode instrumentation for the VEX statement \lstinline[language=vex,mathescape]|t$\langle k\rangle$ = Mul64Fx4(t$\langle i\rangle$, t$\langle j\rangle$)|. Here, the indices $i$, $j$ and $k$ of temporaries for the first factor, second factor, and result of a 256-bit SIMD multiplication of four \lstinline[language=ieee754]|binary64| values are known at instrumentation-time. The (blue) data processed by the original VEX statement, and the (red) shadow data processed by the instrumentation, are samples for values and indices, respectively, that could manifest themselves when the instrumented statement is executed. Only three blocks (in green) are appended to the tape because one of the scalar multiplications involves two passive operands.

\begin{figure}
\centering
\begin{tikzpicture}
\matrix (shad0) at (0,1) [matrix of nodes,nodes={rectangle,thick,draw=black,font=\ttfamily\scriptsize\color{red},minimum width=1.5cm,anchor=center},column sep=-\pgflinewidth,minimum height=0.5cm,row sep=0.1cm] { 0000000a & $*$ & 0000000a & $*$ & 00000000 & $*$ & 00000000 & $*$ \\ 11111134 & $*$ & 2222cafe & $*$ & 00000000 & $*$ & 00000000 & $*$ \\};
\draw ($(shad0-1-1.west)+(-2.8,0)$) node [anchor=west] {\ttfamily t$\langle i + 2m_\text{tmp}\rangle=$};
\draw ($(shad0-2-1.west)+(-2.8,0)$) node [anchor=west] {\ttfamily t$\langle i + m_\text{tmp}\rangle=$};
\matrix (arg0) [matrix of nodes,nodes={rectangle,thick,draw=black,font=\ttfamily\scriptsize\color{blue},minimum width=3.0cm,anchor=center},column sep=-\pgflinewidth,minimum height=0.5cm] { $3.0$ & $0.0$ & $2.0$ & $1.0$  \\};
\draw ($(arg0-1-1.west)+(-2.8,0)$) node [anchor=west] {\ttfamily t$\langle i\rangle=$};
\draw ($(shad0-1-1.west)+(-2.8,0.4)$) node [anchor=south west] {\itshape {\color{blue}Operands} and {\color{red}shadow operands} before and after execution of the instrumented statement:};
\matrix (shad1) at (0,-1.5) [matrix of nodes,nodes={rectangle,thick,draw=black,font=\ttfamily\scriptsize\color{red},minimum width=1.5cm,anchor=center},column sep=-\pgflinewidth,minimum height=0.5cm,row sep=0.1cm] { 000000b0 & $*$ & 000000b1 & $*$ & 00000000 & $*$ & 000000b1 & $*$ \\ 55be5544 & $*$ & 6066beef & $*$ & 00000000 & $*$ & 88081234 & $*$ \\};
\draw ($(shad1-1-1.west)+(-2.8,0)$) node [anchor=west] {\ttfamily t$\langle j + 2m_\text{tmp}\rangle=$};
\draw ($(shad1-2-1.west)+(-2.8,0)$) node [anchor=west] {\ttfamily t$\langle j + m_\text{tmp}\rangle=$};
\matrix (arg1) at (0,-2.5) [matrix of nodes,nodes={rectangle,thick,draw=black,font=\ttfamily\scriptsize\color{blue},minimum width=3.0cm,anchor=center,minimum height=0.5cm},column sep=-\pgflinewidth] { $-4.0$ & $3.7$ & $6.3$ & $2.5$  \\};
\draw ($(arg1-1-1.west)+(-2.8,0)$) node [anchor=west] {\ttfamily t$\langle j\rangle=$};
\matrix (shad2) at (0,-4.3) [matrix of nodes,nodes={rectangle,thick,draw=black,font=\ttfamily\scriptsize\color{red},minimum width=1.5cm,anchor=center},column sep=-\pgflinewidth,minimum height=0.5cm,row sep=0.1cm] { 00000cc0 & $*$ & 00000cc0 & $*$ & 00000000 & $*$ & 00000cc0 & $*$ \\ fedbca03 & $*$ & fedbca04 & $*$ & 00000000 & $*$ & fedbca05 & $*$ \\};
\draw ($(shad2-1-1.west)+(-2.8,0)$) node [anchor=west] {\ttfamily t$\langle k + 2m_\text{tmp}\rangle=$};
\draw ($(shad2-2-1.west)+(-2.8,0)$) node [anchor=west] {\ttfamily t$\langle k + m_\text{tmp}\rangle=$};
\matrix (arg2) at (0,-5.3) [matrix of nodes,nodes={rectangle,thick,draw=black,font=\ttfamily\scriptsize\color{blue},minimum width=3.0cm,anchor=center,minimum height=0.5cm},column sep=-\pgflinewidth] { $-12.0$ & $0.0$ & $12.6$ & $2.5$  \\};
\draw ($(arg2-1-1.west)+(-2.8,0)$) node [anchor=west] {\ttfamily t$\langle k\rangle=$};
\draw ($(shad2-1-1.west)+(-2.8,0.4)$) node [anchor=south west] {\itshape {\color{blue}Left-hand side} and {\color{red}shadow left-hand side} after execution of the instrumented statement:};

\matrix (tape) at (0,-7.9) [matrix of nodes,nodes={rectangle,draw=black,font=\ttfamily\color{green!40!black},minimum width=6cm,minimum height=0.7cm,anchor=center},column sep=-\pgflinewidth,minimum height=0.4cm] { \tapeblock{0000000a\,11111134~~}{~~000000b0\,55be5544}{$-4.0$~}{~$3.0$}  \\  \tapeblock{0000000a\,2222cafe~~}{~~000000b1\,6066beef}{$3.7$~}{~$0.0$} \\ \tapeblock{00000000\,00000000~~}{~~000000b1\,88081234}{$2.5$~}{~$1.0$} \\ };
\fill[black,path fading=north] (tape-1-1.north east) -- +(-\pgflinewidth,0) -- +(-\pgflinewidth,0.5) -- +(0,0.5) -- cycle;
\fill[black,path fading=north] (tape-1-1.north west) -- +(\pgflinewidth,0) -- +(\pgflinewidth,0.5) -- +(0,0.5) -- cycle;
\draw ($(tape-1-1.north)+(0.0,0.4)$) node[font=\color{green!40!black}] {$\vdots~~~~~~~~~~~~~~~~~~~~~~~~\vdots$};
%\draw ($(tape-1-1.west)+(-2.8,0)$) node [anchor=west,font=\ttfamily\scriptsize,align=left] {~~~~~~~~$\vdots$};
\draw ($(tape-1-1.west)+(-2.8,0)$) node [anchor=west,font=\ttfamily\scriptsize,align=left] (indexlabel03) {00000cc0\,fedbca03};
\draw ($(tape-2-1.west)+(-2.8,0)$) node [anchor=west,font=\ttfamily\scriptsize,align=left] (indexlabel04) {00000cc0\,fedbca04};
\draw ($(tape-3-1.west)+(-2.8,0)$) node [anchor=west,font=\ttfamily\scriptsize,align=left] (indexlabel05) {00000cc0\,fedbca05};
\coordinate (tmp1) at ($(indexlabel03.north)+(0,0.4)$);
\coordinate (tmp2) at ($(tape-1-1.north)+(0,0.4)$);
\draw (tmp1|-tmp2) node {$\vdots$};
\draw ($(shad2-1-1.west)+(-2.8,-2.3)$) node [anchor=south west] {\itshape  {\color{green!40!black}Blocks} appended to the tape during execution of the instrumented statement:};
\draw[very thick,<-] ($(tape-3-1.south west)+(-3,0)$) -- ($(tape-1-1.north west)+(-3,0.0)$) node[midway,above,sloped,font=\scriptsize,align=center] {index of \\ block on tape};
\fill[black,path fading=north] ($(tape-1-1.north west)+(-3,0)+(1.5*\pgflinewidth,0)$) -- +(-3*\pgflinewidth,0) -- +(-3*\pgflinewidth,0.5) -- +(0,0.5) -- cycle;
\end{tikzpicture}
\caption{Actions associated to the VEX statement \lstinline[language=vex,mathescape]|t$\langle k\rangle$ = Mul64Fx4(t$\langle i\rangle$, t$\langle j\rangle$)|. Blue numbers are 64-bit floating-point data that the original implementation $f$ processes already. Red integers denote indices, which are stored in shadow memory and processed by the AD instrumentation. Red stars mark unused parts of the shadow memory. The instrumentation also records the green data on the tape. }
\label{fig:vex-multiplication-instrumented}
\end{figure}

\paragraph{Instrumentation of Bitwise Logical Operations, and Limitations} For few arithmetic operations like forming the negative or absolute value, or evaluating an if-then-else construct, compilers frequently emit bitwise logical instead of floating-point instructions. The reverse mode of Derivgrind discovers these ``bit-tricks'' in the same way as the forward mode, and if necessary, appends blocks with the proper indices and partial derivatives to the tape. As we pointed out in \cite{aehle_forward-mode_2022}, the list of possibilities for machine code to ``covertly'' perform real arithmetic through non-floating point instructions is, however, much longer. This has the same consequence in the forward and reverse modes --- lacking a comprehensive strategy to discover unhandled bit-tricks, we have to assume that they do not occur in the program $f$, except for parts of the C math library {\ttfamily libm.so}. Any kind of bit-tricks is allowed in the implementation of C95 {\ttfamily math.h} functions, as we redirect them to wrappers storing analytic derivatives on the tape.

\paragraph{Thoughts on Expression Templates} 
By the \emph{expression templates} approach \cite{hogan_fast_2014}, source code based operator overloading AD tools can operate on entire assignment statements like 
%\begin{lstlisting}[language=C++]
\lstinline|a0 = a1*a2 + a3/a4;|
%\end{lstlisting}
instead of the individual operations. This allows to avoid assigning indices to intermediate results that appear during the evaluation of the right hand side. For instance, CoDiPack records the above statement using four indices (those of \lstinline+a1+, \lstinline+a2+, \lstinline+a3+, \lstinline+a4+) and partial derivatives (\lstinline+a2+, \lstinline+a1+, \lstinline+1.0/a4+, and \lstinline+-a3/(a4*a4)+). The present implementation of Derivgrind sees three binary VEX operations, and thus records three blocks with six indices and partial derivatives in total, unnecessarily assigning an index to either summand. At the moment however, we are not aware of how to discover intermediate variables in the machine code to exploit this kind of optimization.

\paragraph{Storage of the Tape} At the moment, Derivgrind stores the tape in a binary file. Compared to storing the tape on the heap of the Derivgrind process, this has the advantage that a separate program can be used to analyze and evaluate the tape, as we discuss in Section~\ref{sec:tape-evaluation}. The user can still decide to store the tape in RAM by placing the tape file on a RAM disk. 

%In order to support several kinds of bit-tricks, Derivgrind records bitwise logical operations if they could perform real arithmetic, even if they actually do not. Depending on the client program, one could suspect that this causes many unnecessary blocks being written to the tape. Actually however, we confirm that Derivgrind does not record a significant amount of unnecessary operations in our benchmark example in section~\ref{sec:evaluation}.

\subsection{Tape Evaluation}\label{sec:tape-evaluation}
When the client program exits, the recording of the tape is finished. We provide a separate \emph{tape evaluator} program that accesses the tape file, as well as the input and output index files, to perform the following steps:
\begin{enumerate}
\item Determine the number $I$ of blocks on the tape and thus assigned indices, allocate space for $I$ bar values $\bar a_0$, \dots, $\bar a_{I-1}$, and initialize all of them with \lstinline+0.0+.
\item Set the bar values for user-specified indices (typically, all indices in the output index file) to user-specified values.
\item Iterate through the blocks from back to front. When the $i$-th block stores indices $j$ and $\ell$ along with partial derivatives $\tfrac{\partial a_i}{\partial a_j}$ and $\tfrac{\partial a_i}{\partial a_\ell}$, increase $\bar a_j$ by $\tfrac{\partial a_i}{\partial a_j} \cdot \bar a_i$ if $j\ne 0$ and increase $\bar a_\ell$ by $\tfrac{\partial a_i}{\partial a_\ell} \cdot \bar a_i$ if $\ell\ne 0$, according to \eqref{eq:adjoint-update}. Following \eqref{eq:adjoint-zero} is not necessary with a linear index management.
\item Now the bar values for the indices in the input index file contain the sought derivatives.
\end{enumerate}
The tape evaluation program runs independently of Derivgrind and the client program. In principle, it can thus achieve the same performance, measured in processed blocks per time, as other AD tools with a similar tape layout. It is also imaginable to ``import'' the recorded tape into such AD tools. Note that a tape recorded by Derivgrind usually contains more blocks than a tape recorded by a source-code based AD tool using expression templates like CoDiPack.

\paragraph{Record Once, Evaluate Many Times} Certain use cases of AD, like the computation of a Jacobian matrix of a function with several input and output variables, require multiple runs of either forward- or reverse-mode AD, in the following sense.
In reverse-mode AD, the tape must be recorded only once, but then evaluated as many times as there are output variables, each time seeding the bar value of one of them with~$1$ and the other bar values with~$0$. The comparatively large slow-down of the tape recording (see Section~\ref{sec:evaluation}), compared with the tape evaluation, therefore becomes less dominant in this use case. Forward-mode AD must propagate dot values through the client program according to~\eqref{eq:dot-update}, as many times as there are input variables, each time seeding the dot value of one of them with~$1$ and the other dot values with~$0$. As an alternative to the forward-mode feature already present in Derivgrind \cite{aehle_forward-mode_2022}, one could record a tape once and then use the partial derivatives stored on it to repeatedly evaluate~\eqref{eq:dot-update} more quickly.

\section{Identifying Input and Output Variables}\label{sec:identifying-inputs-outputs}
As described in Section~\ref{sec:instrumentation-code}, Derivgrind inserts the index assignment and tape recording logic solely into VEX code, which Valgrind produces from machine code only, without any recourse to the source code of the client. However, Derivgrind also needs to be informed about input variables to treat them as active, and must write indices of input and output index files. In this section, we describe several procedures to specify these variables to Derivgrind. The monitor command and client request interfaces in Section~\ref{sec:io-varname} are analogous to Derivgrind's existing capabilities to enable access to dot values in forward mode, as reported in \cite{aehle_forward-mode_2022}. Section~\ref{sec:io-funarg} describes a new interface tailored to applications in ML.

\subsection{Variables Specified by Line Number and Name in the Source Code}\label{sec:io-varname}
For the scope of this section, we suppose that the Derivgrind user intends to specify input and output variables by references to line numbers and variable names in the source code of the client program. In this case, Derivgrind naturally needs access to those parts of the source code, or debugging symbols generated from it.

\paragraph{Monitor Command Interface} This interface requires the respective parts of the source code to be compiled with debugging symbols and without optimizations. When started with \lstinline[language=bash]|--vgdb-error=0|, Valgrind waits for the user to connect a GDB debugger session. In addition to the usual debugger features like stopping the program execution at a particular line of code and obtaining the memory address of a variable by its name, the resulting ``enhanced'' GDB session allows the user to issue monitor commands, e.\,g.\ in order to read indices and assign new indices to any memory address.

\paragraph{Client Request Interface} This interface requires the user to insert calls to C functions for the input and output variables into the source code of the client program, and recompile (without restrictions on flags). These C functions contain specific sequences of machine code instructions that are recognized by Valgrind, and understood as declarations of AD inputs or outputs. We provide wrappers of these C functions for Fortran and Python, and the precise function names and signatures vary. Generally, an input variable is either passed as a pointer (as in \lstinline[language=bash]|dg_inputf(&var)|), or passed by value and reassigned (as in \lstinline[language=bash]|var=dg_inputf(var)|). Output variables are passed by value.

If the client program has an add-on mechanism exposing the input and output variables to user-supplied code at run-time, client requests can be injected through this mechanism, without modification of the client program's source code at all. As in \cite{aehle_forward-mode_2022}, we provide a Python extension module \lstinline[language=python]|derivgrind.so| that, when dynamically loaded by a Python interpreter, facilitates client requests on the internal C \lstinline[language={}]|double| behind a Python \lstinline[language={}]|float|. 

\subsection{Variables as Arguments to Compiled Functions}\label{sec:io-funarg}
In contrast to Section~\ref{sec:io-varname}, the client code could also be available as a library function with a known signature that comprises all input and output arguments. In this case, the user can apply Derivgrind to a small ``library caller'' program that initializes the input variables, declares every AD input variable with a client request, calls the library function, and declares every AD output variable with a client request. This procedure does not require access to the source code of the client code library.

\paragraph{External Function Wrappers for ML} Similar to \cite{NEURIPS2020_9332c513}, we make this procedure more automatic in the following use case. Suppose that the user needs to include a custom computer-implemented function in a PyTorch or TensorFlow calculation (e.\,g., as a layer in a deep neural network) without reimplementing the function in the respective ML framework. The operator-overloading tools underlying these frameworks allow to temporarily pause the recording of their tape, evaluate an \emph{external function} outside of the ML framework, and specify \emph{custom derivatives} of this external function in terms of additional external code that performs \eqref{eq:adjoint-update}. Derivgrind can be used to automatically provide the custom derivative for a compiled external function.

Listing~\ref{lst:ml-wrapper} illustrates the basic usage of our Python modules \lstinline+derivgrind_torch+ and \lstinline+derivgrind_tensorflow+. In the first block, a simple external function is defined in the C~language, with a specific signature to pass three buffers for non-differentiable parameters (e.\,g.\ hyperparameters), AD inputs (e.\,g.\ information from the previous layer plus trainable parameters of the external function), and AD outputs (e.\,g.\ information for the next layer). The body of this function could call any cross-language or partially closed-source code.

The Python function \lstinline+apply+ in either Python module starts a subprocess to run Derivgrind on a simple library caller program that follows the above steps, and caches the resulting tape and index files. PyTorch's \lstinline+Tensor.backward+ and TensorFlow's \lstinline+GradientTape.gradient+ functions then trigger a subprocess running the tape evaluator on these files to provide the custom gradient.

\begin{lstfloat}
\caption{External function wrappers exposing Derivgrind-differentiated compiled functions in shared objects to the ML frameworks PyTorch and TensorFlow.}
\label{lst:ml-wrapper}
\centering
\vspace{0.2cm}
\begin{minipage}[t]{0.95\textwidth}
\begin{lstlisting}[language=C,frame=single,showlines=true,basicstyle=\footnotesize\ttfamily]
// compilation: gcc thisfile.c -shared -fPIC -o mylib.so -O3
void myfun(int param_size, char* param_buf, 
           int input_count, double* input_buf, 
           int output_count, double* output_buf) {
  output_buf[0] = 3.14*input_buf[0]+5.0+input_buf[1]*input_buf[2];
}
\end{lstlisting}
\end{minipage}
\begin{minipage}[t]{0.95\textwidth}
\begin{lstlisting}[language=python,frame=single,showlines=true,basicstyle=\footnotesize\ttfamily]
import torch
import derivgrind_torch as dg_torch
x = torch.tensor([4.0,-2.0,6.5],dtype=torch.float64, \
                 requires_grad=True)
y = dg_torch.derivgrind("mylib.so","myfun").apply(b"",x,1)
y.backward()
print(*x.grad.numpy()) # -> 3.14 6.5 -2.0
\end{lstlisting}
\end{minipage}
\begin{minipage}[t]{0.95\textwidth}
\begin{lstlisting}[language=python,frame=single,showlines=true,basicstyle=\footnotesize\ttfamily]
import tensorflow as tf
import derivgrind_tensorflow as dg_tf
x = tf.Variable([4.0,-2.0,6.5],dtype=tf.float64)
with tf.GradientTape() as tape:
  y = dg_tf.derivgrind("mylib.so","myfun").apply(b"",x,1)
dy_dx = tape.gradient(y,x)
print(*dy_dx.numpy()) # -> 3.14 6.5 -2.0
\end{lstlisting}
\end{minipage}

\end{lstfloat}

\section{Evaluation}\label{sec:evaluation}
%Validation, compare run-times, tape lengths, ...

\subsection{Regression Tests}\label{sec:regression-tests}
As reported in \cite{aehle_forward-mode_2022}, the existing forward-mode feature of Derivgrind has been validated with an extensive test suite. The suite contains many small client programs, varying the architecture (x86, amd64), the language and compiler (C/\Cpp{} compiled with GCC and Clang, Fortran compiled with GCC), the floating-point format (\lstinline[language=ieee754]|binary32|, \lstinline[language=ieee754]|binary64|, 80-bit x87 type), and the implemented mathematical formula.

For each of the forward-mode testcases, we created a recording-mode testcase that first runs the client program under the recording mode of Derivgrind, and then uses the simple tape evaluator program for both a reverse and a forward evaluation of the tape (Section~\ref{sec:tape-evaluation}).
Except for very few testcases, which \cite{aehle_forward-mode_2022} reports to fail in the forward mode as well due to unhandled bit-tricks, the derivatives of the new recording mode of Derivgrind match the analytic derivatives. 

\subsection{Application to a Python Interpreter} 
Some of the regression tests in Section~\ref{sec:regression-tests} involve Python scripts. In these cases, the actual client program, whose real-arithmetic operations Derivgrind records, is the Python interpreter CPython\footnote{\urldefCPythonHomepage, \urldefCPythonGitHub}, installed on the system by a package manager as a pre-compiled binary. Some of the tests involve the Python C extension module NumPy\footnote{\urldefNumPyHomepage, \urldefNumPyGitHub}, which was partially written in C. Thus, the success of these regression testcases demonstrates that the recording mode of Derivgrind is applicable to a large software project mixing C and Python code, without any changes to the build system. Changes to the source code of CPython were not necessary, as client request functions could be injected with the Python extension module \lstinline[language={}]|derivgrind.so|.

\subsection{Performance Study}
\paragraph{Benchmark} As in \cite{aehle_forward-mode_2022}, we benchmark Derivgrind on a numerical finite-difference solver for the two-dimensional Burgers' partial differential equation (PDE) on an $n_x \times n_x$ grid with $n_t$ time steps \cite{sagebaum_expression_2018,SaAlGauTOMS2019,sagebaum2021index,BluehdornSG2021}. We use client requests to mark all components of the initial state before the first time step as AD inputs, and the norm of the solution after the last time step as the single AD output. 
 The \Cpp{} code merely involves the four elementary arithmetic operations and the square root function. We consider four setups on amd64, combining a compiler (GCC~10.2.1, Clang~11.0.1) and an optimization flag (\lstinline|-O3|, \lstinline|-O0|). All performance tests were executed on an exclusive node with two 64-bit Intel Xeon Gold~6126 processors at \SI{2.6}{\giga\hertz} at the Elwetritsch cluster at TU~Kaiserslautern.  Apart from the measurements presented in the following, we successfully validated the reverse-mode derivatives with the reverse mode of CoDiPack.

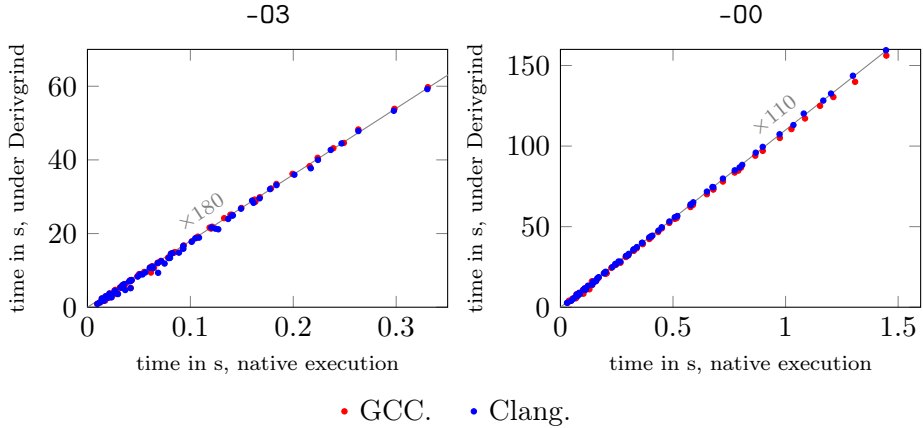
\begin{figure}
\centering
\begin{tikzpicture}
\begin{axis}[title={{\ttfamily -O3}},xlabel={time in s, native execution}, ylabel={time in s, under Derivgrind},xlabel style={align=center,font=\scriptsize},ylabel style={align=center,font=\scriptsize}, tick label style={/pgf/number format/fixed}, width=0.43\textwidth, height=5cm, xmin=0,xmax=0.35,ymin=0,ymax=70]
\addplot[red,mark=*,only marks,mark size=1pt] table[x index=0, y index=1] {bar_time/bar_time_amd64_g++_o3};
%\addplot[red,mark=square,only marks,mark size=1pt] table[x index=0, y index=1] {bar_time/bar_time_x86_g++_o3};
\addplot[blue,mark=*,only marks,mark size=1pt] table[x index=0, y index=1] {bar_time/bar_time_amd64_clang++_o3};
%\addplot[blue,mark=square,only marks,mark size=1pt] table[x index=0, y index=1] {bar_time/bar_time_x86_clang++_o3};

\draw [gray,thin] (axis cs: 0,0) -- (axis cs: 1,180) node[pos=0.12,sloped,above] {\scriptsize $\times 180$};
\end{axis}
\end{tikzpicture}
\begin{tikzpicture}
\begin{axis}[title={{\ttfamily -O0}},xlabel={time in s, native execution}, ylabel={time in s, under Derivgrind},xlabel style={align=center,font=\scriptsize},ylabel style={align=center,font=\scriptsize}, tick label style={/pgf/number format/fixed}, width=0.43\textwidth, height=5cm, xmin=0,xmax=1.6,ymin=0,ymax=160]
\addplot[red,mark=*,only marks,mark size=1pt] table[x index=0, y index=1] {bar_time/bar_time_amd64_g++_o0};
\addplot[blue,mark=*,only marks,mark size=1pt] table[x index=0, y index=1] {bar_time/bar_time_amd64_clang++_o0};
\draw [gray,thin] (axis cs: 0,0) -- (axis cs: 2,220) node[pos=0.50,sloped,above] {\scriptsize $\times 110$};
\end{axis}
\end{tikzpicture}
\hfill \\
    \raisebox{0.20em}{\addlegendimageintext{red,mark=*,only marks,mark size=1pt}} GCC. \quad
 %   \raisebox{0.20em}{\addlegendimageintext{red,mark=*,only marks,mark size=0.5pt}} GCC, tape in RAM. \quad
     \raisebox{0.20em}{\addlegendimageintext{blue,mark=*,only marks,mark size=1pt}} Clang. \quad
%     \raisebox{0.20em}{\addlegendimageintext{blue,mark=*,only marks,mark size=0.5pt}} Clang, tape in RAM. \quad
\caption{Effect of Derivgrind's recording mode on the run-time of the Burgers benchmark.}
\label{fig:burgers-time}
%\Description{Two scatter plots titled -O3, -O0, with x-axis label ``time in s, native execution'' and y-axis label ``time in s, under Derivgrind''.}
\end{figure}

\paragraph{Run-Time of the Tape Recording} Figure~\ref{fig:burgers-time} displays the effect of Derivgrind's recording mode on the run-time, each marker representing a problem instance with $n_x=100,120,\dots,400$ and $n_t=100, 200, 300, 400$. Run-time measurements, both under Derivgrind's recording mode and in native execution without Derivgrind, were performed by the client program, calculating the system time difference before and after solving the PDE. This excludes Derivgrind's constant start-up time of up to a few seconds. Measurements were repeated 25 (\lstinline|-O3|) or 5 (\lstinline|-O0|) times, and averaged. 
The tape file was placed in the ramdisk directory \lstinline|/dev/shm|.

Asymptotically, Derivgrind scales the run-time of the client program by a proportionality factor of about~180 for optimized and~110 for unoptimized builds.

% TIME

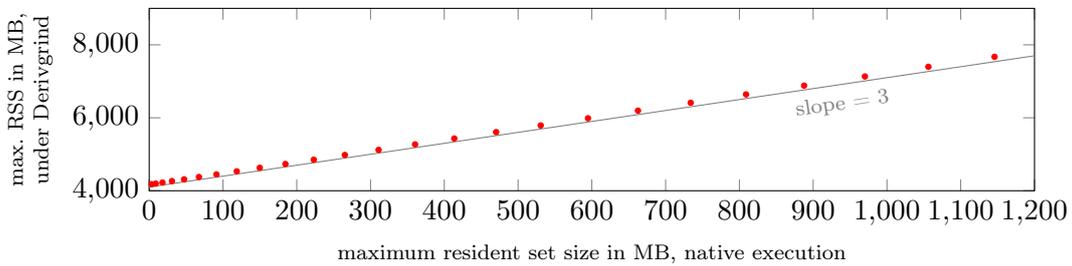
\begin{figure}
\centering
\begin{tikzpicture}
\begin{axis}[title={},xlabel={maximum resident set size in MB, native execution}, ylabel={max.\ RSS in MB, \\ under Derivgrind},xlabel style={align=center,font=\scriptsize},ylabel style={align=center,font=\scriptsize}, width=0.9\textwidth, height=4cm, xmin=0,xmax=1200,ymin=4000,ymax=9000]
\addplot[red,mark=*,only marks,mark size=1pt] table[x expr=\thisrowno{0}/1024.0, y expr=\thisrowno{1}/1024.0] {bar_mem/bar_mem_amd64_g++_o3};
\draw [gray,thin] (axis cs: 0,4100) -- (axis cs: 1200,7700)  node[pos=0.78,sloped,below] {\scriptsize slope~=~3};
\end{axis}
\end{tikzpicture}
\caption{Effect of Derivgrind's recording mode on the maximum resident set size of the Burgers benchmark, in a setup using GCC and \lstinline|-O3|. RAM or disk space occupied by the tape file is not included here.} 
\label{fig:burgers-mem}
%\Description{Scatter plot with x-axis label ``maximum resident set size in MB, native execution'' and y-axis label ``max. RSS in MB, under Derivgrind''.}
\end{figure}

\paragraph{Memory Complexity of the Tape Recording, Excluding Tape} Figure~\ref{fig:burgers-mem} displays the effect of Derivgrind on the basic memory consumption of the process, in terms of the maximum resident set size (RSS) measured by the GNU \lstinline|time| command. The RSS adds up allocations made by the client code, the Valgrind core, and the Derivgrind tool including the shadow memory tool, but excludes any RAM space required to store the tape file on the ramdisk. We considered problem instances with $n_x=200, 400, \dots, 5000$ and $n_t=4$. Only the results for GCC with \lstinline|-O3| are shown, as changing to Clang and/or \lstinline|-O0| had only minor effects.  

The slope of~3 is consistent with the fact that two bytes of shadow memory are allocated for every byte of memory that the client uses. The instance-independent constant allocation of about \SI{4.1}{\giga\byte} on amd64 is mainly occupied by the shadow memory tool. Configuring it differently, we can decrease the constant allocation to below \SI{0.1}{\giga\byte}, at the price of slightly increasing the run-time.

%% MEMORY

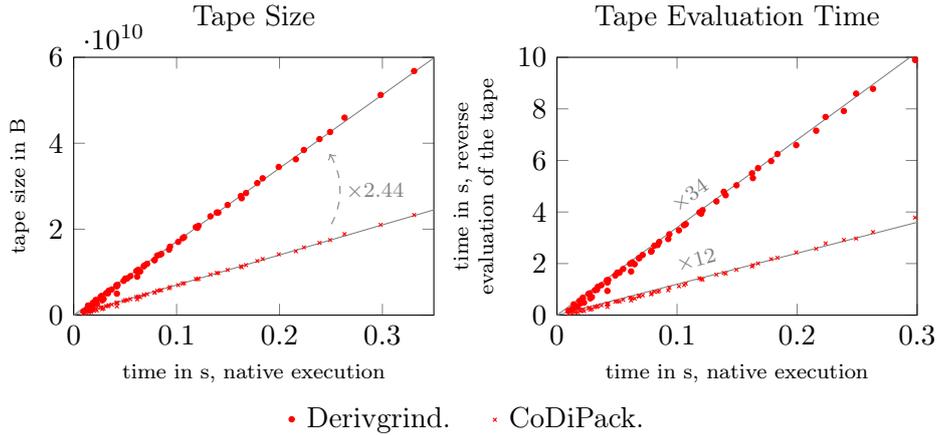
\begin{figure}
\centering
\begin{tikzpicture}
\begin{axis}[title={Tape Size},xlabel={time in s, native execution}, ylabel={tape size in \si{\byte}},xlabel style={align=center,font=\scriptsize},ylabel style={align=center,font=\scriptsize}, tick label style={/pgf/number format/fixed}, width=0.43\textwidth, height=5cm, xmin=0,xmax=0.35,ymin=0,ymax=60e9]
\addplot[red,mark=*,only marks,mark size=1pt] table[x index=0, y index=2] {bar_tapesize/bar_tapesize_amd64_g++_o3_5to1merge};
%\addplot[blue,mark=*,only marks,mark size=1pt] table[x index=0, y index=2] {bar_tapesize/bar_tapesize_amd64_clang++_o3};
\addplot[red,mark=x,only marks,mark size=1pt] table[x index=0, y index=1] {bar_tapesize/bar_tapesize_amd64_g++_o3_5to1merge};
%\addplot[blue,mark=x,only marks,mark size=1pt] table[x index=0, y index=1] {bar_tapesize/bar_tapesize_amd64_clang++_o3};
\draw [gray,thin] (axis cs: 0,0) -- (axis cs: 1,171*1000*1000*1000) node[pos=0.24,sloped] (dg-line) {};
\draw [gray,thin] (axis cs: 0,0) -- (axis cs: 1,70*1000*1000*1000) node[pos=0.24,sloped] (codi-line) {};
\draw[gray,->,thin,dashed,bend right,shorten <=0.1cm, shorten >=0.1cm] (codi-line) to (dg-line) node[midway,right] {};
\draw[gray] ($0.5*(codi-line)+0.5*(dg-line)$) node[right] {\scriptsize ~~$\times 2.44$};
\end{axis}
\end{tikzpicture}
%%%%%%%%%%%%%%%%%%%%%%%%%%%%%%%%%%%%%
\begin{tikzpicture}
\begin{axis}[title={Tape Evaluation Time},xlabel={time in s, native execution}, ylabel={time in s, reverse \\ evaluation of the tape},xlabel style={align=center,font=\scriptsize},ylabel style={align=center,font=\scriptsize}, tick label style={/pgf/number format/fixed}, width=0.43\textwidth, height=5cm, xmin=0,xmax=0.3,ymin=0,ymax=10]
\addplot[red,mark=*,only marks,mark size=1pt] table[x index=0, y index=2] {bar_timeeval/bar_timeeval_amd64_g++_o3};
%\addplot[blue,mark=*,only marks,mark size=1pt] table[x index=0, y index=2] {bar_timeeval/bar_timeeval_amd64_clang++_o3};
\addplot[red,mark=x,only marks,mark size=1pt] table[x index=0, y index=1] {bar_timeeval/bar_timeeval_amd64_g++_o3};
%\addplot[blue,mark=x,only marks,mark size=1pt] table[x index=0, y index=1] {bar_timeeval/bar_timeeval_amd64_clang++_o3};
\draw [gray,thin] (axis cs: 0,0) -- (axis cs: 1,12) node[pos=0.12,sloped,above] {\scriptsize $\times 12$};
\draw [gray,thin] (axis cs: 0,0) -- (axis cs: 1,34) node[pos=0.12,sloped,above] {\scriptsize $\times 34$};
\end{axis}
\end{tikzpicture}
\hfill \\
    \raisebox{0.20em}{\addlegendimageintext{red,mark=*,only marks,mark size=1pt}} Derivgrind. \quad
    \raisebox{0.20em}{\addlegendimageintext{red,mark=x,only marks,mark size=1pt}} CoDiPack. \quad
\caption{Tape sizes and reverse tape evaluation run-times for Derivgrind and CoDiPack, plotted against the native run-time of the client program built with GCC and \lstinline+-O3+.}
\label{fig:burgers-timeeval}
\end{figure}

\paragraph{Tape Size} The left part of Figure~\ref{fig:burgers-timeeval} shows that the ratio between the sizes of the tapes recorded by Derivgrind and CoDiPack, for one run of the Burgers' PDE benchmark compiled with GCC and \lstinline+-O3+, is around $2.44$ across all problem instances. Likewise, the numbers of pairs of non-zero indices and partial derivatives on the respective tapes consistently follow a ratio close to $1.66$. The dominant part of the PDE solver's real arithmetic are four \Cpp{} statements in the body of a nested loop, which Derivgrind and CoDiPack represent with \SI{448}{\byte} vs.\ \SI{184}{\byte} of tape space ($\tfrac{\SI{448}{\byte}}{\SI{184}{\byte}} = 2.43\ldots$) and 25 vs.\ 15 pairs of non-zero indices and partial derivatives ($\tfrac{25}{15} = 1.66\ldots$), respectively. CoDiPack needs less tape space and less pairs because it can make use of expression templates, and allows for a more flexible tape layout. The ratios could be even larger if the right-hand sides in the \Cpp{} code were more complex. This result also tells us that Derivgrind does not record a significant amount of unnecessary operations on amd64. These observations also hold when the Clang compiler and/or \lstinline+-O0+ is used.

% REVERSE EVALUATION TIME

\paragraph{Run-Time of the Tape Evaluation} 
As the reverse tape evaluation procedure is simple and unrelated to the machine code of the client program, we can expect its run-time performance to catch up with the state of the art, in which the run-time is memory bandwidth bound \cite{BluehdornSG2021,TOWARA201519}.
The right side of Figure~\ref{fig:burgers-timeeval} compares the reverse tape evaluation run-times of Derivgrind and CoDiPack for the same set of problem instances considered in the recording mode run-time measurements. We observe that Derivgrind's tape evaluation procedure, taking about $2.8$ times longer than CoDiPack's, is not much worse given that the tape is $2.4$ times longer. 

The Burgers' benchmark case has a high proportion of floating-point operations involving active variables. In codes that perform more non-floating-point or passive operations, we expect a smaller tape size to be recorded per second of native run-time. Then, the ratio between the reverse tape evaluation run-time and the native run-time drops below the value of~34 found in Figure~\ref{fig:burgers-timeeval}. When the full AD workflow involves a single tape recording but many tape evaluations, as discussed in Section~\ref{sec:tape-evaluation}, this lower factor dominates the full AD run-time.

\paragraph{Performance Study on x86} On x86, we observed much higher run-time scaling factors up to about 1500 and similar memory scaling factors of about~3. The preceding statements about tape length and correctness apply to small problem instances on x86 as well. We could not test large instances because CoDiPack stores the tape in the memory of the client process, which is limited to about \SI{3}{\giga\byte} on x86.

\section{Conclusion}\label{sec:conclusion}
In this work, we have presented a methodology to compute reverse-mode automatic derivatives for compiled programs. 

When started with the novel option \lstinline[mathescape]+--record=$\langle\text{\rmfamily \emph{path}}\rangle$+, the dynamic binary instrumentation tool Derivgrind records the real-arithmetic evaluation tree of the client program in a tape file. 
To a large extent, the fundamental procedures are analogous to the existing forward-mode capability of Derivgrind already presented in \cite{aehle_forward-mode_2022}. Floating-point values, and possibly other data, are identified with integer indices in the shadow storage. The instrumentation of real-arithmetic operations records these indices and the respective partial derivatives on the tape, and assigns new indices. We defined recording-mode instrumentation for the same set of VEX statements, expressions and operations as in the forward mode. Thus, Derivgrind operates under the same assumption that the client code performs all real arithmetic through the respective floating-point instructions, or a specific set of bit-tricks. Unhandled bit-tricks are allowed in the C math library, as we adapted Derivgrind's math function wrappers to push analytic derivatives to the tape.

The index handling and tape recording process is fully independent of the source code of the client program, and thus applicable to cross-language or partially closed-source software projects. Only in order to specify input and output variables by their variable names and lines of code, access to the respective sections of the source code is, naturally, required. To this end, we have defined monitor commands and client requests similar to those in the forward mode.

The price for Derivgrind's general scope of application, compared to state-of-the-art high-performance operator overloading tools like CoDiPack, is its higher demand for run-time and memory. In the tape recording phase, Derivgrind slows down our benchmark client program by a factor of about 110 or 180 for unoptimized and optimized builds on amd64, respectively. The size of the recorded tape was larger because the tape layout is simpler, and because statement-wise preaccumulation using expression templates is (currently) not feasible.

Our software package includes a simple tape evaluator that propagates bar values through the reversed evaluation tree stored in the tape file, from the specified output variables back to the specified input variables, given their respective indices. It is also imaginable to ``import'' the tape into other AD tools. We have successfully validated the derivatives for many small regression test samples, the Python interpreter CPython, and a PDE solver.

ML researchers can use Derivgrind very easily via the external function interfaces of PyTorch and TensorFlow, to include compiled code in ML workflows. We believe that machine code based AD unlocks many opportunities to explore optimization and ML techniques in new application domains.

\section*{Acknowledgements}
Max Aehle gratefully acknowledges funding from the research training group SIVERT by the German federal state of Rhineland-Palatinate.

We are grateful to the authors of Valgrind for creating such a highly versatile framework. Karl Cronburg's shadow memory library\footnote{\url{https://github.com/cronburg/shadow-memory}} was very helpful in the initial phase of the development of Derivgrind.

\bibliography{lib}
\bibliographystyle{unsrt}

\end{document}

%% file: lstinit.tex
\lstdefinelanguage{vex}{
    keywords = {}
}
\lstdefinelanguage{ieee754}{
    keywords = {}
}
\lstdefinelanguage{asm}{ % inline assembler
    keywords = {}
}

%% file: main.bbl
\begin{thebibliography}{10}

\bibitem{hascoet_tapenade_2013}
Laurent Hascoet and Valérie Pascual.
\newblock {The Tapenade automatic differentiation tool: Principles, model, and
  specification}.
\newblock {\em {ACM} Trans. Math. Softw.}, 39(3):1--43, 2013.

\bibitem{NEURIPS2020_9332c513}
William Moses and Valentin Churavy.
\newblock {Instead of Rewriting Foreign Code for Machine Learning,
  Automatically Synthesize Fast Gradients}.
\newblock In H.~Larochelle, M.~Ranzato, R.~Hadsell, M.~F. Balcan, and H.~Lin,
  editors, {\em Advances in Neural Information Processing Systems}, volume~33,
  pages 12472--12485. Curran Associates, Inc., 2020.

\bibitem{Walther2012Gsw}
Andrea Walther and Andreas Griewank.
\newblock {Getting started with {ADOL-C}}.
\newblock In Uwe Naumann and Olaf Schenk, editors, {\em Combinatorial
  Scientific Computing}, chapter~7, pages 181--202. Chapman-Hall CRC
  Computational Science, 2012.

\bibitem{SaAlGauTOMS2019}
Max Sagebaum, Tim Albring, and Nicolas~R. Gauger.
\newblock {High-Performance Derivative Computations using {CoDiPack}}.
\newblock {\em ACM Transactions on Mathematical Software (TOMS)}, 45(4), 2019.

\bibitem{maclaurin2015autograd}
Dougal Maclaurin, David Duvenaud, and Ryan~P Adams.
\newblock {Autograd: Effortless Gradients in Numpy}.
\newblock In {\em ICML 2015 AutoML Workshop}, volume 238, page~5, 2015.

\bibitem{NEURIPS2019_9015}
Adam Paszke, Sam Gross, Francisco Massa, Adam Lerer, James Bradbury, Gregory
  Chanan, Trevor Killeen, Zeming Lin, Natalia Gimelshein, Luca Antiga, Alban
  Desmaison, Andreas Kopf, Edward Yang, Zachary DeVito, Martin Raison, Alykhan
  Tejani, Sasank Chilamkurthy, Benoit Steiner, Lu~Fang, Junjie Bai, and Soumith
  Chintala.
\newblock {PyTorch: An Imperative Style, High-Performance Deep Learning
  Library}.
\newblock In H.~Wallach, H.~Larochelle, A.~Beygelzimer, F.~d\textquotesingle
  Alch\'{e}-Buc, E.~Fox, and R.~Garnett, editors, {\em Advances in Neural
  Information Processing Systems 32}, pages 8024--8035. Curran Associates,
  Inc., 2019.

\bibitem{tensorflow2015-whitepaper}
Mart\'{i}n Abadi, Ashish Agarwal, Paul Barham, Eugene Brevdo, Zhifeng Chen,
  Craig Citro, Greg~S. Corrado, Andy Davis, Jeffrey Dean, Matthieu Devin,
  Sanjay Ghemawat, Ian Goodfellow, Andrew Harp, Geoffrey Irving, Michael Isard,
  Yangqing Jia, Rafal Jozefowicz, Lukasz Kaiser, Manjunath Kudlur, Josh
  Levenberg, Dandelion Man\'{e}, Rajat Monga, Sherry Moore, Derek Murray, Chris
  Olah, Mike Schuster, Jonathon Shlens, Benoit Steiner, Ilya Sutskever, Kunal
  Talwar, Paul Tucker, Vincent Vanhoucke, Vijay Vasudevan, Fernanda Vi\'{e}gas,
  Oriol Vinyals, Pete Warden, Martin Wattenberg, Martin Wicke, Yuan Yu, and
  Xiaoqiang Zheng.
\newblock {TensorFlow}: Large-scale machine learning on heterogeneous systems,
  2015.
\newblock Software available from tensorflow.org.

\bibitem{aehle_forward-mode_2022}
Max Aehle, Johannes Blühdorn, Max Sagebaum, and Nicolas~R. Gauger.
\newblock {Forward-Mode Automatic Differentiation of Compiled Programs}.
\newblock Preprint, arXiv:2102.11572.

\bibitem{valgrind-paper}
Nicholas Nethercote and Julian Seward.
\newblock {Valgrind: A Framework for Heavyweight Dynamic Binary
  Instrumentation}.
\newblock {\em SIGPLAN Not.}, 42(6):89–100, jun 2007.

\bibitem{valgrind-doc}
Julian Seward, Nicholas Nethercote, Tom Hughes, Jeremy Fitzhardinge, Josef
  Weidendorfer, et~al.
\newblock {Valgrind Documentation, Release 3.19.0, 11 Apr 2022}, 2022.

\bibitem{sagebaum2021index}
Max Sagebaum, Johannes Blühdorn, and Nicolas~R. Gauger.
\newblock {Index handling and assign optimization for Algorithmic
  Differentiation reuse index managers}.
\newblock arXiv cs.MS 2006.12992, 2021.

\bibitem{shadowmem-nethercote}
Nicholas Nethercote and Julian Seward.
\newblock {How to Shadow Every Byte of Memory Used by a Program}.
\newblock In {\em Proceedings of the 3rd International Conference on Virtual
  Execution Environments}, VEE '07, page 65–74, New York, NY, USA, 2007.
  Association for Computing Machinery.

\bibitem{ieee754}
{IEEE 754}.
\newblock {IEEE Standard for Floating-Point Arithmetic}.
\newblock Technical report, IEEE, 2008.

\bibitem{hogan_fast_2014}
Robin~J. Hogan.
\newblock {Fast Reverse-Mode Automatic Differentiation Using Expression
  Templates in C++}.
\newblock {\em ACM Trans. Math. Softw.}, 40(4), jul 2014.

\bibitem{sagebaum_expression_2018}
Max Sagebaum, Tim Albring, and Nicolas~R. Gauger.
\newblock {Expression templates for primal value taping in the reverse mode of
  algorithmic differentiation}.
\newblock {\em Optimization Methods and Software}, 33(4):1207--1231, 2018.

\bibitem{BluehdornSG2021}
Johannes Bl\"{u}hdorn, Max Sagebaum, and Nicolas~R. Gauger.
\newblock {Event-Based Automatic Differentiation of OpenMP with OpDiLib}.
\newblock {\em ACM Trans. Math. Softw.}, nov 2022.
\newblock Just Accepted.

\bibitem{TOWARA201519}
Markus Towara, Michel Schanen, and Uwe Naumann.
\newblock {MPI-Parallel Discrete Adjoint OpenFOAM}.
\newblock {\em Procedia Computer Science}, 51:19--28, 2015.
\newblock International Conference On Computational Science, ICCS 2015.

\end{thebibliography}
